# Structural regulation of mechanical gating in molecular junctions

B. Pabi[1], J. Šebesta [2,3], R. Korytár[2], O. Tal[4] and A. N. Pal[1,4]

1. Department of Condensed Matter and Materials Physics, S. N. Bose National Centre for Basic Sciences, Sector III, Block JD, Salt Lake, Kolkata700106, India

2. Department of Condensed Matter Physics, Faculty of Mathematics and Physics, Charles University, CZ-121 16 Praha 2, Czech Republic

3. Materials Theory, Department of Physics and Astronomy, Uppsala University Box 516, 751 20 Uppsala, Sweden

4. Department of Chemical and Biological Physics, Weizmann Institute of Science, Rehovot 7610001, Israel

**Abstract:**

In contrast to silicon-based transistors, single molecule junctions can be gated by simple mechanical means. Specifically, charge can be transferred between the junction's electrodes and its molecular bridge when the interelectrode distance is modified, leading to variations in the electronic transport properties of the junction. While this effect has been studied extensively, the influence of the molecule orientation on mechanical gating has not been addressed, despite its potential influence on the gating effectiveness. Here, we show that the same molecular junction can experience either clear mechanical gating or none, depending on the molecule orientation in the junctions. The effect is found in silver-ferrocene-silver break junctions, and analyzed in view of ab-initio and transport calculations, where the influence of molecular orbitals geometry on charge transfer to or from the molecule is revealed. The molecule orientation is thus a new degree of freedom that can be used to optimize mechanically-gated molecular junctions.

**Main Text**

One of the fascinating properties of molecular junctions is their ability to function as nanoscale electro-mechanical devices. In particular, single-molecule junctions allow the study of coupling between mechanical and electronic degrees of freedom in a structure of a typical single-nanometer size that has a dominant quantum nature, and a pronounced orbital character. This combination has been used to study diverse phenomena, including electron-phonon interaction[1–6], quantum interference[7,8] and charge reorganization[9] in the miniaturization limit for electronic conductors. Interestingly, in addition to the more standard electrostatic gating of molecular junctions, these junctions can be mechanically gated. By changing the interelectrode distance in the electrode-molecule-electrode junction, molecular energy levels can be shifted to a lower or higher energy and charge can be transferred from the electrodes to the molecule or vice versa. Consequentially, the electronic transport characteristics of the junction may change. Mechanically-gated molecular junctions have been extensively studied both experimentally and theoretically[7–15], for example in the context of nanoscale image charge[9] and optimization of thermoelectricity[14]. However, the influence of the molecule orientation on mechanical gating has not been examined. Such influence can be an attractive route for optimization and regulation of mechanical gating with implications on charge, spin, and heat transport in molecular junctions. Here, we show that mechanical gating of molecular junctions can be dramatically affected by the molecule orientation, where the same molecular junction experiences either a clear mechanical gating or the absence of such an effect, depending on the orientation of the molecule with respect to the electrodes. By comparing experiments and calculations, this behavior can be related to the orbital nature at the metal-molecule interfaces, allowing the identification of the necessary conditions for mechanical gating. The reported findings in this letter show that the orientation of the molecule is an important factor for the design of mechanically-gated molecule junctions, and emphasizes the importance of orbital orientation in the general process of metal-molecule charge transfer.

We study molecular junctions based on suspended individual ferrocene molecules between two silver (Ag) electrode tips. A break junction setup is used to form in-situ these molecular junctions (Figure 1) in a cryogenic environment (~ 4.2K). The ferrocene molecules are introduced from a local heated molecular source into a cold atomic-scale Ag junction during repeated junction breaking and making cycles (see Section 1 in the Supporting Information). Measurements of current as a function of applied voltage across the junctions (I-V curves) for different junction realizations reveal two distinctive cases, denoted here as type 1 and type 2, (Figures 2a, b). The presented I-V curves were measured following a repeated reduction in the interelectrode separation. Several steps can be observed in the I-V curves, translated in Figures 2c, d to peaks in the corresponding differential conductance versus voltage (dI/dV-V) curves. As will be further discussed below with the aid of ab-initio calculations, the peaks originate from the contributions of molecular orbitals to the conductance. Therefore, shifts in the voltage at which the peaks are observed correspond to shifts in the molecular energy levels with respect to the Fermi level of the electrodes. Figure 2c, reveals that the peaks of type 1 are shifted to a lower voltage when the interelectrode separation is reduced. Namely, the molecular level or levels that dominate transport in type 1 are shifted towards the Fermi level when the junction is squeezed. In contrast, in Figure 2d, the peaks for type 2 are not shifted in response to a similar mechanical manipulation. Specifically, the inset of Figure 2c, presents a significant shift from 1.365 V to 0.675 V for a reduction in the interelectrode distance of ~ 0.6 Å for type 1, whereas a similar behavior is not seen in the inset of Figure 2d for type 2. This is an indication for mechanical induced molecular energy shifts, or mechanical gating, in type 1 and the absence of this effect in type 2 (see Supporting Information for similar type 1 and 2 behaviors in different realizations of the Ag-ferrocene-Ag molecular junctions).

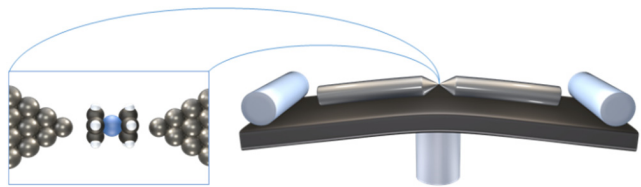

**Figure 1.** Break junction setup. Schematics of the used break junction setup, in which the distance between the Ag electrode tips can be adjusted in sub-Ångstrom resolution, and an illustration of a Ag-ferrocene-Ag single-molecule junction.

The same effect can also be presented using transition voltage spectroscopy (TVS) plots, where the I-V characteristics in Figure 2a, b are re-plotted in Figures 2e, f in terms of $ln(I/V^2)$ as a function of $1/|V|$ [16–20]. These plots are expected to have a minimum at a certain transition voltage, $V_{trans}$, whenever the current as a function of voltage evolves from a linear dependence to more than a quadratic dependence[19]. We note that for metallocene molecules with direct contacts to the electrodes (namely, no anchoring side groups are used) the transition voltage dependence on the energetic difference between the electrodes' Fermi level and the closest molecular energy level(s) is to date unknown. However, regardless the exact dependence, shifts in the transition voltage serve as an indication for shifts in the energy of the molecular levels that dominate electron transport and/or systematic variations in their coupling to the continuum states of the electrodes (i.e., electrode-molecules coupling)[21,22]. As seen below using calculations, we expect dominant level shifts and moderate variations in the electrode-molecule coupling. This is manifested as significant shifts in the calculated transmission peaks with generally modest variations in the peak widths. The TVS plots in Figures 2e, f, show a clear characteristic minima, with a corresponding transition voltage. The response of the transition voltage to changes in the interelectrode distance is observed in the insets of Figures 2e, f, where a shift is seen for type 1 but not for type 2 (see Supporting Information for similar data for other junction realizations). We can take advantage of the presence of peaks in the dI/dV that correspond to molecular levels to shed light on the relation between transition voltage shifts and molecular level shifts. When the interelectrode distance is reduced by 0.6 Å, the transition voltage shifts by 370 mV. Assuming for simplicity a symmetric voltage drop on each electrode-molecule contact, which can be justified as a crude approximation by the roughly symmetric locations of the positive and negative peaks in Figure 2c), the 690 mV shift in the examined peak location in Figure 2c corresponds to a shift of 690/2=345 mV of the molecular levels that dominate the electron transport. Note that for a symmetric voltage drop, the location of the transporting molecular levels with respect to the electrode Fermi level is given by half the voltage at which a peak in the dI/dV curve appears (this is illustrated below with the aid of calculated transmissions and dI/dV curves). We can conclude that the observed mechanical gating leads to shifts in the transition voltage that are rather similar in magnitude to the roughly estimated shifts in the molecular energy levels (this can also be seen in Supplementary Figure S2). Our findings therefore show that TVS is a good indicator for level shifts in the examined junction.

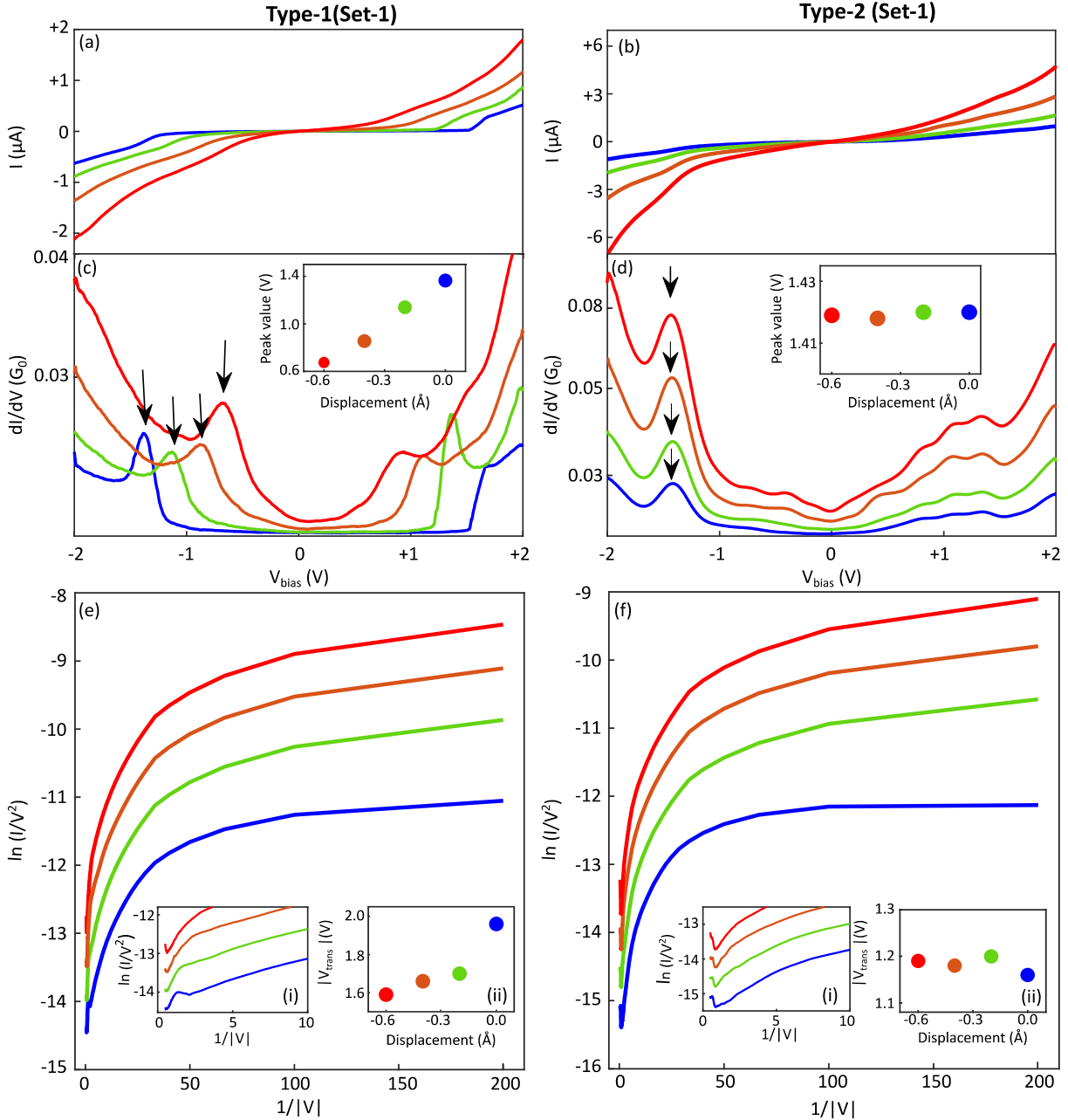

**Figure 2.** Current-voltage, differential conductance spectra and transition voltage spectroscopy (TVS) plots. (a,b) Four spectra of current as a function of voltage measured at different interelectrode displacements (for color code and displacement see c,d Insets) in Ag-ferrocene molecular junctions with (a, type 1) and without (b, type 2) mechanical gating response. (c) Differential conductance as a function of applied voltage for the junction studied in a. (d) Same as c but with data collected for the molecular junction studied in b. Insets (c,d) Absolute values of peak position (marked with arrows in c,d) as a function of interelectrode displacement. (e-f) TVS plots constructed from the same I-V spectra presented in a,b, showing $\ln(I/V^2)$ versus $1/|V|$ for spectra with (e, type 1) and without (f, type 2) mechanical gating response. For consistency, the negative side of the I-V curves is considered for TVS analysis. Insets (i): Zoomed view of the TVS plots to better present the change of transition voltage upon squeezing. Insets (ii): Transition voltage (absolute values) as a function of interelectrode displacement for type 1 and 2.

To better understand the nature of type 1 and 2, we turn to density functional theory (DFT) and electron transport calculations (see Supporting Information for details), as presented in Figures 3a-f. For the range of interelectrode displacements that is considered in the experiments the molecular junction can adapt, according to our calculations, two distinct stable configurations with parallel and perpendicular molecular orientations with respect to the electrode axis, as illustrated in the insets of Figures 3a, b. The calculated total energy as a function of interelectrode separation is presented in Figure 3f for each one of these junction configurations. The two energy curves have a clear minimum at different interelectrode separations. At short distances, the perpendicular configuration is energetically preferred, while for longer distances the parallel configuration is more stable. A similar behavior was also reported for Ag-vanadocene-Ag junctions by Pal et al.[23]

Figures 3a, b, provide the calculated transmission for various interelectrode distances for the two configurations. The transmissions for the parallel configuration can essentially be understood as being rigidly shifted toward lower energies upon mechanical squeezing. Namely, a mechanical gating is observed[13]. Below the Fermi energy (taken as zero) pairs of narrow resonances can be seen, while above the Fermi energy a single broad resonance is found. Its peak transmission is 2, pointing to a two-channel or two-orbital origin. The transmissions for the perpendicular configuration have a richer structure, with narrow peaks on both sides of the Fermi energy. Also, the evolution of transmission as a function of stretching is more complex in this case than for the parallel molecular configuration. However, in contrast to the former case, no clear shifts in the transmission peaks are observed when the interelectrode distance is modified in the given range. Thus, mechanical gating is not found for the perpendicular configuration. We note that in both configurations the transmission resonances are highly asymmetric due to quantum interference[24], similarly to the asymmetries reported for molecular junctions based on a ferrocene derivative[8].

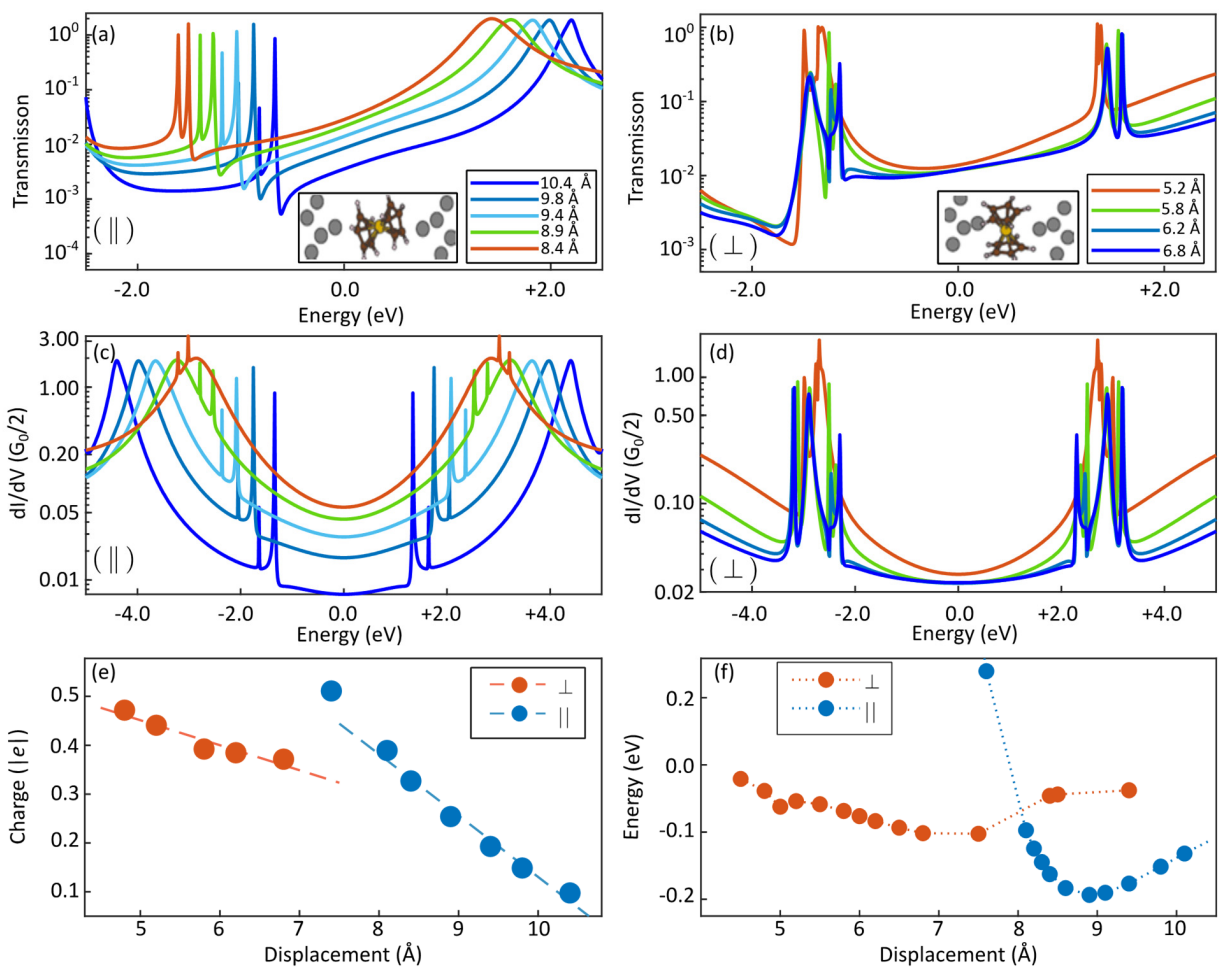

**Figure 3.** Transport calculations. (a,b) Calculated transmission for parallel (a) and perpendicular (b) molecule orientations in the junction at a varying distance between the electrode tips. Insets: ball-and-stick models of the calculated structures (only a small part of the electrodes is shown). (c,d) Differential conductance of the parallel (c) and perpendicular (d) configurations at the same varying electrode tip distances as in a and b. (e) Charging of the ferrocene molecule in the parallel (blue) and perpendicular configurations (orange). (f) Total energy as a function of interelectrode displacement. (⊥, ∥) denote perpendicular and parallel molecular orientation with respect to the long junction axis.

To have a more transparent comparison with the measured data, the described transmission can be directly converted to calculated differential conductance, as presented in Figures 3c, d, assuming a similar voltage drop across the two electrode-molecule contacts (see Eq. S3 in the Supporting Information). When the molecule is oriented in parallel to the junction axis, mechanical gating is seen, as was found in the experiments for type 1. However, for the perpendicular molecular orientation a similar effect cannot be found, in agreement with the absence of mechanical gating in the measurements of type 2. Note that the features in Figure 3c, d are sharper than found in the measured spectra in Figure 2c, d, since the experimental data is widened by the finite temperature and mainly by the lock-in modulation used for differential conductance measurements.

Figure 3e reveals charge transferred from the electrodes to the molecule in equilibrium as a function of inter electrode displacement. The slope is larger in the parallel junction, suggesting that charge reorganization plays an important role in the gating mechanism. The mechanical gating response for the parallel molecule configuration and the absence of this effect for the perpendicular configuration can be ascribed to the orientation of the molecular orbitals that dominate transport with respect to the electrodes, and their coupling to the frontier electrode states. Figure 4 presents calculated iso-surfaces of the two degenerate lowest unoccupied molecular orbitals (LUMOs) for an isolated ferrocene, as well as the LUMOs for the parallel and perpendicular junction configurations (calculations were done for a cluster of a single molecule bridging two metal apices, as explained in the Supporting Information). As can be seen, the junction LUMOs, which are associated with the broad transmission peak above the Fermi energy in the parallel configuration, and the two narrow peaks above the Fermi energy in the perpendicular configuration, resemble the LUMOs of the isolated molecule. These orbitals have a significant $\pi$-character on the carbon rings, *i.e.* the wave function spreads away from the ring plane perpendicularly[25]. The remaining contribution to the LUMOs comes from the Fe 3*d* atomic orbitals. Therefore, the molecular LUMOs overlap with the electrodes much more efficiently in the parallel configuration than in the perpendicular configuration. This difference makes the LUMO coupling to the frontier electrode states in the parallel configuration more sensitive to changes in the interelectrode distance (seen in Figure 3a as a different width of the broad LUMO peak for different interelectrode distance) than in the perpendicular case. Thus, the orientation of the LUMO with respect to the electrodes affects the mechanical gating efficiency: When the electrodes are pointing to the less-localized part of the LUMO on the carbon ring, mechanical manipulations likely induce orbital modifications and associated charge transfer. In contrast, when the electrodes are pointing towards the more localized part of the LUMO on the Fe ion, mechanical manipulation has a reduced effect on the local orbital structure and the associated charge redistribution. These findings illustrate that molecular orientation, as well as the distribution of molecular orbitals in space are important parameters for the design and control of mechanical gating in molecular junctions and generally for mechanically-induced charge transfer in metal-molecule interfaces. Interestingly, based on our analysis, efficient mechanical gating is promoted by delocalization of transporting molecular orbitals that point towards the electrodes, while for electrostatic gating in single molecule transistors, efficient gating by a gate electrode is often promoted by more localized orbitals on the molecular bridge, with a rather low coupling to the electrode states.

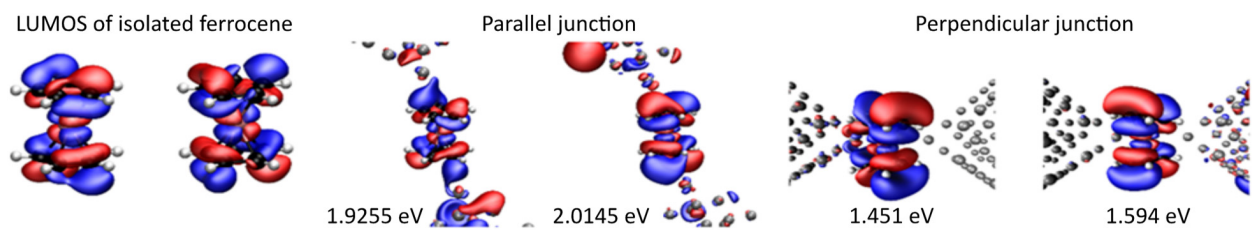

**Figure 4.** Isosurfaces of the calculated orbitals that dominate electronic transport. Left: Isosurface plots of the two (degenerate) LUMOs of an isolated ferrocene. Center and right, respectively: isosurface plots of selected electron wavefunctions of the Ag-ferrocene-Ag junction and their energies (with respect to Fermi energy) for the parallel and perpendicular configurations (at interelectrode separations of 9.81Å and 6.2 Å, which correspond to the blue curves in Figures 3a,b). These energies lie in the immediate vicinity of the unoccupied transmission resonances. All isosurfaces contain 93% of the wavefunction. The plots also contain ball and stick models of the structures (color coding of the atoms: white (H), black (C), pink (Fe) and silver (Ag)).

To conclude, we showed by experiments and calculations that mechanical gating of molecular junctions depends on the orientation of the molecule in the junction. In the extreme demonstrated case, the same molecular junction either experience mechanical gating or not, depending on the molecule orientation with respect to the electrodes, as well as on the nature of the interaction between the molecular orbitals and the continuum states of the electrodes. These findings emphasize the importance of geometry and local orbital structure in the context of charge transfer across metal-molecule interfaces, and point towards a way to control mechanical gating of charge, spin, and heat transport in molecular junctions.

# Structural regulation of mechanical gating in molecular junctions

B. Pabi[1], J. Šebesta[2,3], R. Korytár[2], O. Tal[4] and A. N. Pal[1,4]

1. Department of Condensed Matter and Materials Physics, S. N. Bose National Centre for Basic Sciences, Sector III, Block JD, Salt Lake, Kolkata700106, India

2. Department of Condensed Matter Physics, Faculty of Mathematics and Physics, Charles University, CZ-121 16 Praha 2, Czech Republic

3. Materials Theory, Department of Physics and Astronomy, Uppsala University Box 516, 751 20 Uppsala, Sweden

4. Department of Chemical and Biological Physics, Weizmann Institute of Science, Rehovot 7610001, Israel

**Content:**
1. Experimental methods and conductance characterization of Ag and Ag-ferrocene junction.
2. Additional experimental data.
3. Ab initio calculations of structure and electronic transport.

# 1. Experimental methods and conductance characterization of Ag and Ag-ferrocene junction.

A mechanically controllable break junction set up is used to form molecular junction. A Ag wire (99.997%, 0.1 mm, Alfa Aesar) with a notch at its middle is fixed on top of a flexible substrate (1mm thick phosphor bonze covered by 100μm Kapton foil). This structure is placed in a vacuum chamber that is cooled by liquid helium to ~ 4.2K. Using a three-point bending mechanism, the substrate is pushed and bent. This process stretches the wire notch, resulting a gradual reduction in its cross section down to the atomic scale. Fine bending of the substrate is achieved by a piezo electric actuator (PI P- 882 PICMA) which is driven by a 24-bit DAQ card (PCI 4461- National Instruments), followed by a piezo driver (SVR 150/1, Piezomechanik). Sufficient bending of the substrate results wire breaking into two wire segments with freshly exposed atomic tips that are formed in an ultraclean cryogenic environment. These sharp wire segments serve as electrodes. Molecular junctions are prepared by continuously breaking and reforming a metallic atomic contact between the electrode tips, while sublimating ferrocene molecules (99.5%, Alpha Aesar, further purified in situ) from a locally heated molecular source towards the atomic contact. Direct current (d.c.) conductance is measured when the junction is gradually pulled apart to form conductance versus distance traces. A constant 500 mV from the DAQ card is supplied to a divider by 10 (to improve signal to noise ratio) and the resulted 50 mv bias is applied across the junction. The resulted current output from the junction is amplified by a I/V preamplifier (Femto DLPCA-200), and the amplified signal is recorded by the same DAQ card. The d.c. conductance of the junction is thus the measured current divided by the applied voltage (50 mV). Inter-electrode displacement is estimated based on the dependence of the tunneling current on the electrode separation, following a standard process[1]. Differential conductance spectra (dI/dV versus V) is recorded using a standard lock-in technique. A reference sine signal from a lock-in amplifier (SR830) with a peak to peak voltage of 10 mv and a frequency of ~3.333 kHz is added to a d.c. voltage and the total voltage is divided by 10 to improve signal to noise ratio. The response alternating current (a.c.) is probed by a lock-in amplifier (SR830) and recorded by the DAQ card The differential conductance spectra are obtained by dividing the alternating current signal (dI) with the applied alternating voltage bias (dV), as a function of a swiped d.c. voltage bias (V).

The d.c. conductance of the junction (current/voltage) is recorded during repeated junction stretching as a function of relative interelectrode displacement. First, a bare Ag junction was characterized. Figure S1a presents in blue examples for conductance traces as a function of interelectrode displacement. During the elongation of the Ag junction, the conductance decreases in steps when the contact diameter is reduced. The last conductance plateau at ~1 $G_0$ ($G_0 = 2e^2/h$, is the conductance quantum, where $e$ is the electron charge and $h$ is the Plank's constant) provides the conductance of a single atom contact between the Ag electrode tips[2]. After the insertion of ferrocene molecules, tilted plateaus below the ~1 $G_0$ step are clearly seen (Figure S1a, red traces). To statistically characterize the most probable conductance features, conductance histograms (Figure S1b) are constructed from 5,000 and 10,000 consecutive conductance-displacement traces for Ag junctions before (blue) and after (red) the introduction of ferrocene, respectively. While the conductance histogram for bare Ag junctions reveals a peak at ~1 $G_0$ (Inset of Figure S1b) that is associated with the most probable conductance of single Ag atom contacts, after the introduction of ferrocene an additional conductance peak beneath the $1G_0$ peak is observed. Gaussian fitting of the corresponding peak, shown by a black dash dot line in Figure S1b yields a most probable conductance value of $(7.25 \pm 0.06) \times 10^{-3}$ $G_0$ for Ag-ferrocene molecular junctions.

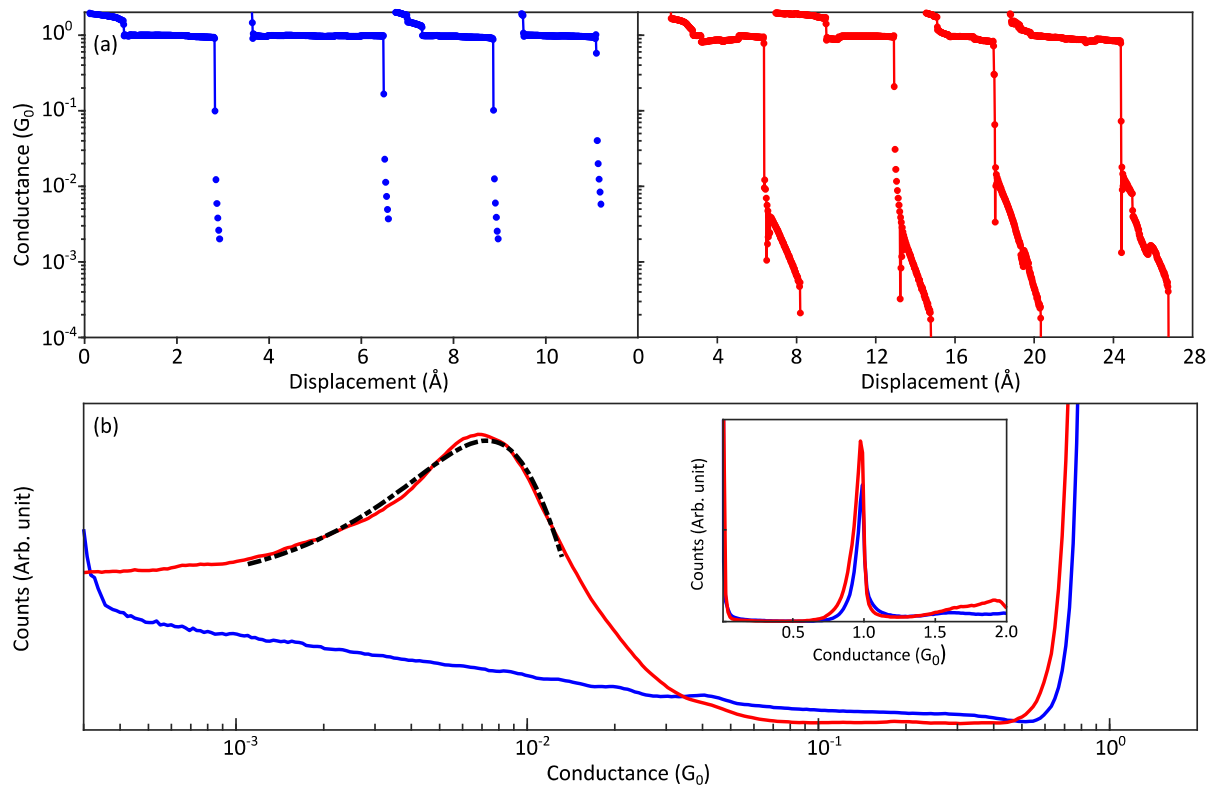

**Figure S1.** Conductance characterization of Ag and Ag-ferrocene junction. (a) Left and right panel: Traces of conductance vs. interelectrode displacement recorded during the breaking of Ag atomic-scale junctions before (blue) and after (red) the insertion of ferrocene molecules, recorded at 50 mV bias voltage. Traces are shifted horizontally for clarity. (b) Conductance histogram for Ag (blue) and Ag-ferrocene (red) junctions, prepared from 5,000 and 10,000 consecutive conductance traces using 80 bins per decade. Black dash dot line represents a Gaussian fitting to the observed peak. Inset: Conductance histogram of the same junctions, constructed from 300 bins in a linear scale.

## 2. Additional experimental data.

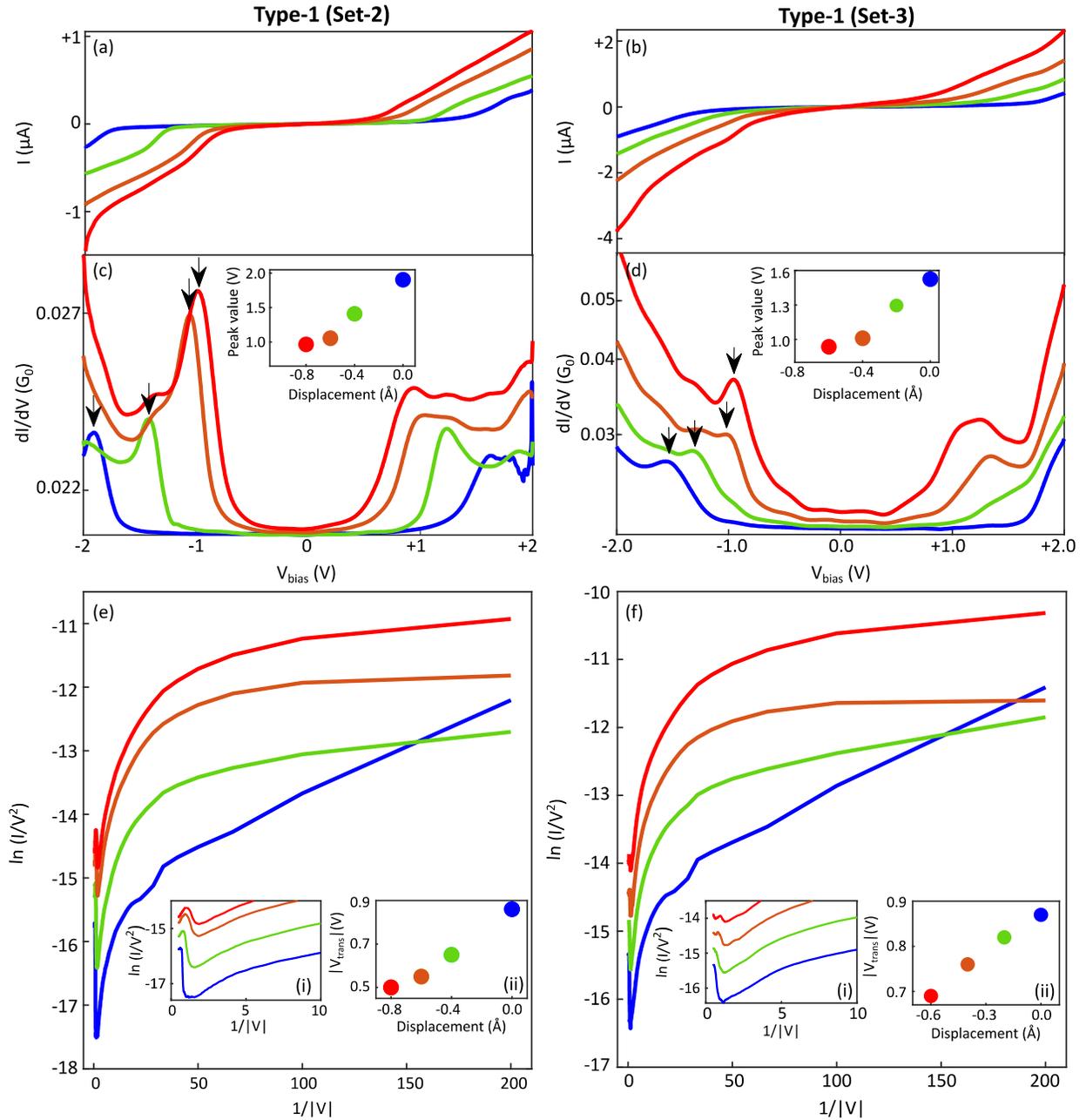

**Figure S2.** Current-voltage curves, differential conductance spectra, and TVS plots for type 1. (a,b) Current vs. voltage measured at different interelectrode displacements in Ag-ferrocene junctions with mechanical gating response (type 1). (c) Differential conductance vs. voltage for the junction studied in a. (d) Same as c but with data collected for the molecular junction studied in b. Insets (c,d) Absolute values of peak position (marked with arrows in c,d) vs. interelectrode displacement. (e-f) TVS plots constructed from the same I-V spectra presented in a,b. For consistency, the negative side of the I-V curves is considered for TVS analysis. Insets (i): Zoomed view of the TVS plots. Insets (ii): Transition voltage (absolute values) vs. interelectrode displacement.

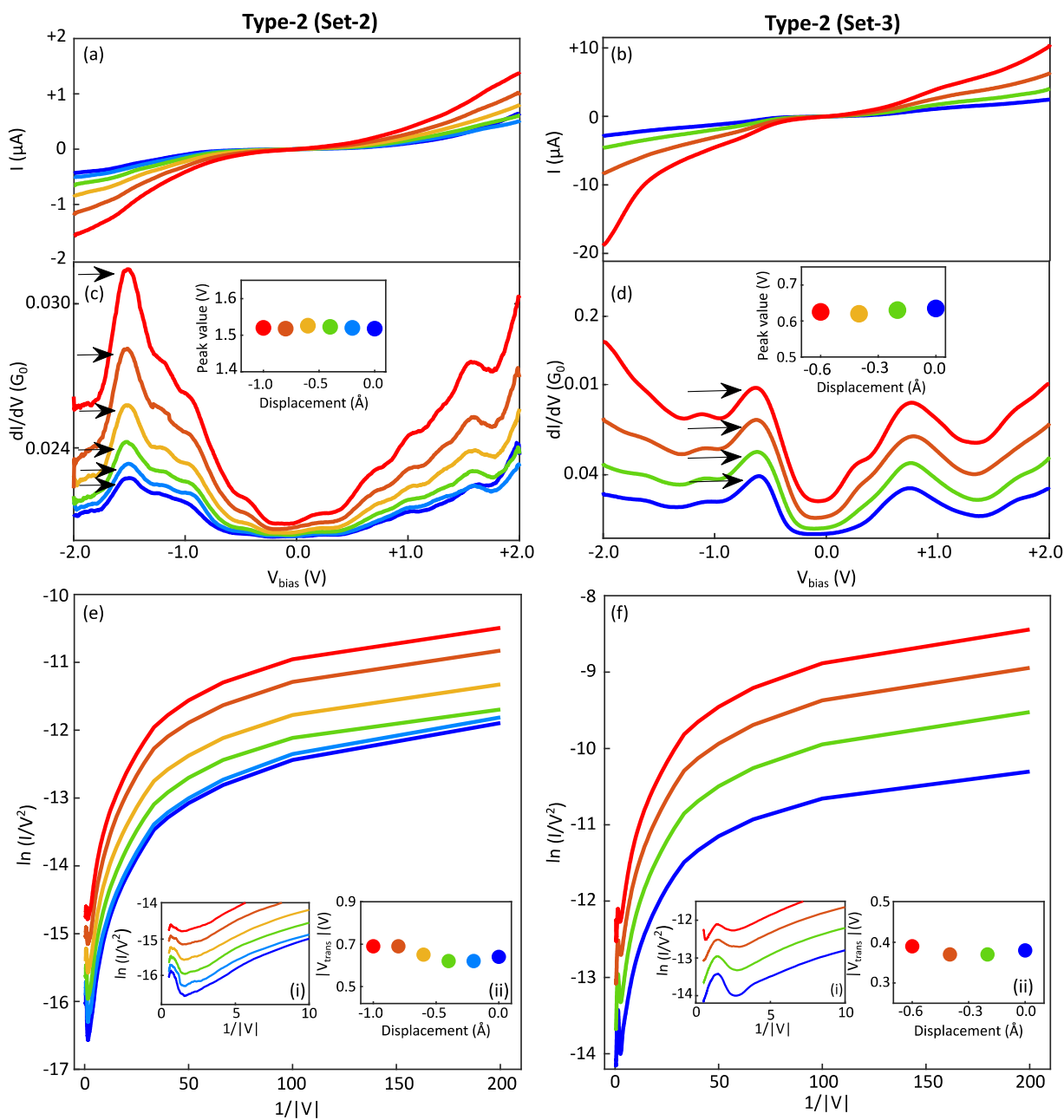

**Figure S3.** Current-voltage curves, differential conductance spectra, and TVS plots for type 2. (a,b) Current vs. voltage measured at different interelectrode displacements in Ag-ferrocene junctions with no mechanical gating response (type 2). (c) Differential conductance vs. voltage for the junction studied in a. (d) Same as c but with data collected for the molecular junction studied in b. Insets (c,d) Absolute values of peak position (marked with arrows in c,d) vs. interelectrode displacement. (e-f) TVS plots constructed from the same I-V spectra presented in a,b. Insets (i): Zoomed view of the TVS plots. Insets (ii): Transition voltage (absolute values) vs. interelectrode displacement.

## 3. Ab initio calculations of structure and electronic transport.

Theoretical transmission functions were obtained by ab-initio transport calculations based on the non-equilibrium Green's function (NEGF) technique and effective scattering states given by the Kohn-Sham density functional theory (DFT), see Ref. 3 for a comprehensive overview. Utilizing the TURBOMOLE package[4,5], electronic wave-functions were described through Gaussian type orbitals of the def2-TZVP basis set[6,7], while the exchange-correlation functional of the Perdew-Burke-Ernzerhof (PBE) type was employed[8].

**Structures and structural relaxation**

Prior the transport calculation, possible molecule's binding conformations to the electrodes were studied through molecule's structure optimization by the quasi-Newton-Raphson method[9]. All the examined structures were non-periodic clusters composed of the ferrocene molecule and an attached pair of Ag electrodes with a pyramidal shape, as schematically depicted in the insets of Figure 2 and in Figure S4.

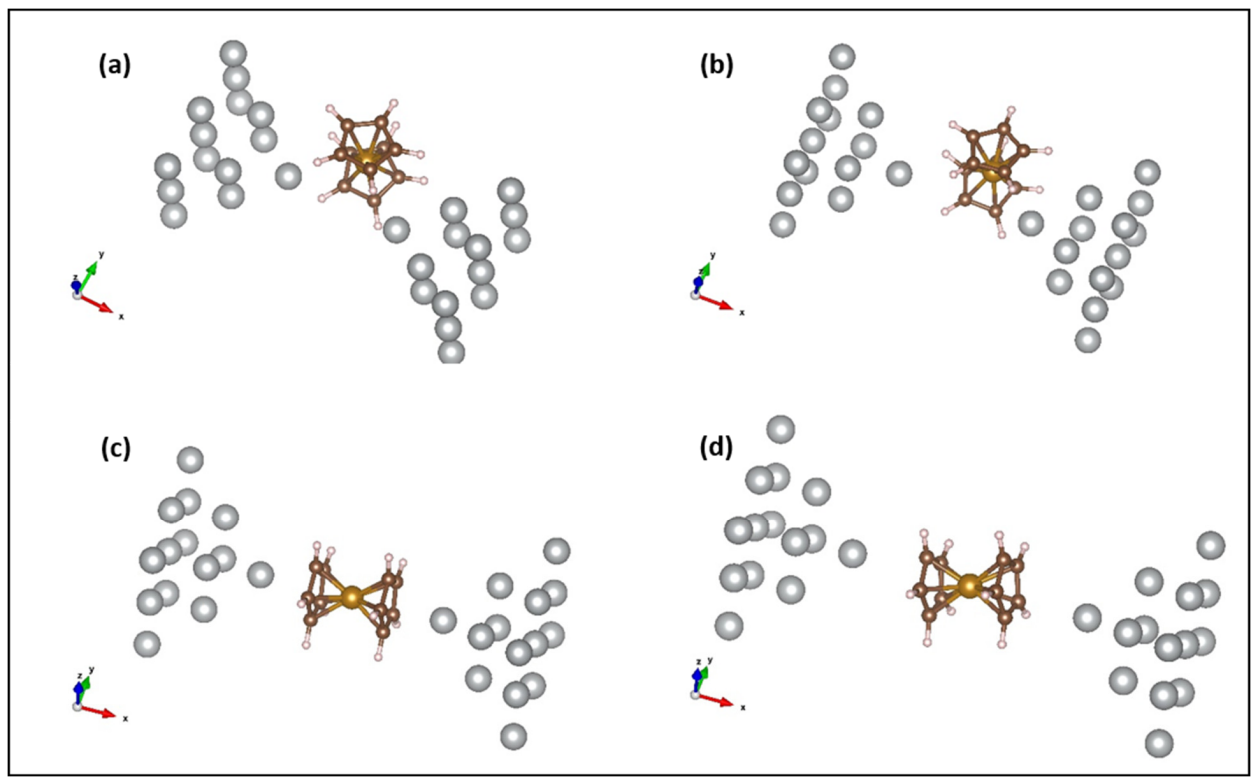

**Figure S4.** Relaxed structures. Parallel molecule orientation with respect to the electrode axis: **(a)** d=5.2 Å **(b)** d=6.2Å. Perpendicular molecule orientation: **(c)** d=8.1 Å **(d)** d=9.8Å.

The Ag electrodes were cut out of a face-centered Ag crystal along the (100) direction considering a bulk lattice parameter[10]. They consist of an apex atom and a set of atomic layers with a growing number of atoms. During relaxation, we kept the Ag atomic positions fixed. The sequence of pyramids was necessary to study of the conductance's convergence when the size of the pyramidal electrode tips increases. Due to computational demands, the structure optimization was performed for clusters with three-layered leads and the molecule coordinates were pre-relaxed in the gas phase.

**Quantum transport**

Differential conductance was calculated via NEGF formalism[3] employing the AITRANSS package[11–13]. The infinite electrodes were accounted for by self-energy operators applied to the Ag pyramids following Ref. 13. At a given voltage V, the electric current is given by the expression

$$I(V) = \frac{2e}{h} \int_{-\infty}^{\infty} T(E,V) [f_L(E,V) - f_R(E,V)] dE \quad \text{(S1)}$$

where the factor 2 accounts for spin degeneracy, $T(E,V)$ is the transmission function and $f_{L,R}$ denotes the Fermi-Dirac distributions of the left (L) and right (R) electrodes. We assumed that the external bias voltage causes a symmetric shift of the left and right chemical potentials, $\mu_{LR} = \mu \pm \frac{eV}{2}$, where μ stands for the equilibrium chemical potential. Further, we adopted the zero-temperature form of $f_{L,R}$; this approximation is valid as long the transmission function $T(E,V)$ varies slowly within the energy range $\in (\mu_{L,R} - k_B T, \mu_{L,R} + k_B T)$, where $k_B T$ is the working temperature multiplied by the Boltzmann constant. Considering the expression $f_{L,R}(E,V) = \theta\left(\mu \pm \frac{eV}{2} - E\right)$, it follows that the current and its derivative are given by

$$I(V) = \frac{2e}{h} \int_{E_F - \frac{eV}{2}}^{E_F + \frac{eV}{2}} T(E,V) \, dE \quad \text{(S2)}$$

$$\frac{dI}{dV} = \frac{2e^2}{h} \frac{1}{2} \left[ T\left(E_F + \frac{eV}{2}\right) + T\left(E_F + \frac{eV}{2}\right) \right] \quad \text{(S3)}$$

where the Fermi energy $E_F$ equals the chemical potential at $T = 0$. We neglected the explicit voltage dependence in the transmission, which can quantitatively, and sometimes qualitatively change the differential conductance. The bias voltage can lead to non-equilibrium Stark effect of the resonances[11]. This effect can simply be understood as orbital modifications due to the presence of an effective electric field induced by the voltage $V$. Consequently, the coupling of an orbital to the two electrodes is more asymmetric at $V \neq 0$ than at V = 0 and the heights of the corresponding transmission peaks diminish. Therefore, the theoretical differential conductance of type-1 conformation could overemphasize the resonance heights and the inclusion of non-equilibrium Stark effect would diminish them, making them more consistent with the experimental data.